\documentclass[twocolumn,showpacs,prb]{revtex4}
\usepackage{graphicx}

\begin{document}

\preprint{}
\title{
Magnetization process in a chiral $p$-wave superconductor 
with multi-domains 
}

\author{Masanori Ichioka} 
\email{oka@mp.okayama-u.ac.jp}
\author{Yasushi Matsunaga} 
\author{Kazushige Machida}
\affiliation{Department of Physics, Okayama University,
         Okayama 700-8530, Japan}

\date{\today}

\begin{abstract}
A simulation study for the magnetization process 
is performed for the multi-domain state in a chiral $p$-wave superconductor, 
using the time-dependent Ginzburg-Landau theory. 
The external field penetrates inside as core-less vortices 
through the domain wall, forming the vortex sheet structure. 
We find that, with increasing magnetic fields, 
the domain walls move so that the unstable domains shrink to vanish. 
Therefore, the single domain structure is realized at higher fields. 
\end{abstract}

\pacs{74.25.Qt, 74.20.Rp, 74.20.De, 74.70.Pq}


\maketitle 

A quasi-two dimensional superconductor ${\rm Sr_2RuO_4}$ is an 
unconventional superconductor, and the pairing symmetry is suggested 
to be a chiral $p$-wave pairing with the basic form 
$p_\pm=p_x \pm i p_y$ and 
in-plane equal-spin pairing.\cite{Maeno,Rice} 
The spin-triplet pairing is supported by the fact that 
the Knight shift does not change below the superconducting 
transition temperature $T_{\rm c}$.\cite{Ishida} 
Since the spontaneous moment appears below $T_{\rm c}$ 
in the observation of the muon spin relaxation ($\mu$SR), 
time-reversal symmetry of 
the pairing function is broken.\cite{Luke} 
The internal field distribution of the vortex state, 
which is observed by a small angle neutron scattering (SANS), 
is consistent to the chiral $p$-wave pairing state.\cite{Kealey}  
While the pairing function of ${\rm Sr_2RuO_4}$ may have 
an additional horizontal node, 
such as $(p_x \pm i p_y)\cos p_z$,\cite{Hasegawa}  
it is intrinsic in our study that the pairing function has a factor 
of the chiral component  $p_x \pm i p_y$. 

In the chiral $p$-wave superconductor, 
a $p_+$ state and a $p_-$ state are degenerate in free energy. 
Therefore, the multi-domain structure may realize, i.e., 
some regions in a sample are $p_+$ domains and others are $p_-$ domains. 
Between the $p_+$ and $p_-$ domains, 
domain walls appear as topological defects, 
which are not easily destroyed.~\cite{MatsumotoD,Sigrist}  
In the vortex state when a magnetic field ${\bf H}$ is 
applied to $\hat{z}$ direction, 
the degeneracy of the $p_+$ and $p_-$ states is removed. 
When ${\bf H} \parallel \hat{z}$ and charge $e>0$ 
(or equivalently when ${\bf H} \parallel -\hat{z}$ and $e<0$), 
the free energy of the $p_-$ state is lower than that of 
the $p_+$ state,~\cite{IchiokaP} 
because the vortex structures are different between 
the $p_+$ and $p_-$ domains. 
In the $p_-$ ($p_+$) state, the opposite 
$p_+$ ($p_-$) component induced around the vortex core has large (small) 
amplitude.~\cite{IchiokaP,Heeb,Takigawa}  
Since the $p_-$ state is stable in the vortex state, 
it is interesting to see how the multi-domain structure at a zero field 
changes to the single $p_-$ domain, when a magnetic field is applied. 
We note that the chirality dependence is defined relative to the 
magnetic field direction, the free energy of the $p_+$ state becomes smaller 
when the magnetic field is applied to the reverse direction $-z$. 
Study of vortices trapped at the domain wall is also important 
since they are considered to have 
a strange structure called a ``vortex sheet'',\cite{Parts} 
where half flux-quantum vortices are aligned along the domain 
wall.\cite{Sigrist,Matsunaga,Matsunaga2,BabaevL,BabaevB}

The purpose of this paper is to study the magnetization process 
in order to understand how the multi-domain structure 
changes to the single $p_-$ domain by applying magnetic fields  
in the chiral $p$-wave superconductor. 
We also investigate the roles and properties of the vortex sheet structure 
at the domain wall appearing in the magnetization process. 
The static properties of the vortex or domain wall structure were 
studied also by the two-component Ginzburg-Landau (GL) 
theory.\cite{MatsumotoD,Sigrist,Heeb,Agterberg,Kita} 
For the study of the magnetization process, we use the 
time-dependent Ginzburg-Landau (TDGL) theory 
as a phenomenological approach in our qualitative study, 
with expectation that vortices move 
so as to approach the free energy minimum state. 
The magnetization process of a conventional superconductor was studied by 
Kato {\it et al.}\cite{KatoEnomotoC,KatoEnomoto}
The present authors studied the static and dynamical properties of 
the vortex sheet structure in a two-component superconductor 
based on the TDGL theory.\cite{Matsunaga,Matsunaga2}
In this paper, we use the GL free energy for the chiral $p$-wave 
superconductor, and study the magnetization process. 

To obtain the two-component GL equation 
in the chiral $p$-wave superconductor, 
the pair potential is decomposed as  
$\Delta({\bf r},{\bf p})
=\eta_+({\bf r})\phi_+({\bf p})
+\eta_-({\bf r})\phi_-({\bf p})$  
with the order parameter $\eta_\pm({\bf r})$, 
where ${\bf r}$ is the center-of-mass coordinate of the Cooper pair.  
The pairing functions $\phi_\pm({\bf p})$, 
depending on the relative momentum ${\bf p}$ of the pair, 
are given by the chiral $p$-wave type such as $p_x \pm i p_y$. 
The GL free energy density is written as 
\begin{eqnarray} && 
\tilde{f} 
=-\left(1-\frac{T}{T_{{\rm c}}}\right)
  \left( |\eta_+|^2 + |\eta_-|^2 \right)
+\frac{1}{2} |\eta_+|^4 
+\frac{1}{2} |\eta_-|^4 
\nonumber \\ && 
+2 |\eta_+|^2  |\eta_-|^2  
+C_1 \left( 
\eta_-^{\ast 2} \eta_+^2 + \eta_+^{\ast 2} \eta_-^2  \right)
+ \eta_+^\ast (q_x^2+q_y^2) \eta_+
\nonumber \\ && 
+ \eta_-^\ast (q_x^2+q_y^2) \eta_- 
+ C_2 \left( 
\eta_+^\ast q_-^2 \eta_- +\eta_-^\ast q_+^2 \eta_+ \right) 
\nonumber \\ && 
+ C_3 \left( 
\eta_+^\ast q_+^2 \eta_- +\eta_-^\ast q_-^2 \eta_+ \right) 
\label{eq:fe}
\end{eqnarray} 
in the dimensionless form,~\cite{Matsunaga}   
where $q_\pm=q_x \pm i q_y$, 
${\bf q}=(\hbar/ i)\nabla-2\pi(2e/hc){\bf A}$ 
with vector potential ${\bf A}$. 
$hc/2|e|=\phi_0$ is a flux-quantum. 
The coefficients are related to the pairing function and the Fermi 
surface structure as  
\begin{eqnarray} &&
C_1= \frac{  \langle \phi_-^{\ast 2} \phi_+^2 \rangle }
          {2 \langle |\phi_+|^4 \rangle } , \qquad 
C_2= \frac{  \langle v_+^2 \phi_+^\ast \phi_- \rangle }
          {2 \langle v_+ v_- |\phi_+|^2 \rangle } , \
\nonumber \\ && 
C_3= \frac{  \langle v_-^2 \phi_+^\ast \phi_- \rangle }
          {2 \langle v_+ v_- |\phi_+|^2 \rangle } , 
\label{eq:factor}
\end{eqnarray}
where $v_\pm=(v_x \pm i v_y)/2$ with a Fermi velocity $(v_x,v_y)$, 
and $\langle \cdots \rangle$ indicates the average on ${\bf p}$ 
along the Fermi surface. 
$C_1$ and $C_3$ are anisotropy parameters which are finite 
when the pairing functions or the Fermi surface have fourfold symmetric 
structure. 
For the isotropic case, $C_1=C_3=0$ due to 
the vanishing Fermi surface average in  Eq. (\ref{eq:factor}).
As the detailed forms of $\phi_\pm$ and the Fermi surface structure 
have not been established yet, 
we treat the coefficients in Eq. (\ref{eq:factor}) 
as arbitrary parameters. 
In this study we set $C_1=C_3=0$ for simplicity to exclude 
the additional anisotropy effect.  
We show results calculated for $C_2=0.3$ and $T=0.5 T_c$.

In our simulations, we use the TDGL equation coupled with Maxwell 
equations,~\cite{KatoEnomotoC,KatoEnomoto} 
\begin{eqnarray} && 
\frac{\partial}{\partial t}\eta_1 
= -\frac{1}{12} \frac{\partial \tilde{f}}{\partial \eta_1^\ast}, 
\qquad 
\frac{\partial}{\partial t}\eta_2 
= -\frac{1}{12} \frac{\partial \tilde{f}}{\partial \eta_2^\ast}, 
\label{eq:TDGL}
\\ && 
\frac{\partial}{\partial t}{\bf A} =  \tilde{\bf j}_{\rm s} 
-\kappa^2 \nabla\times{\bf B}, 
\qquad 
{\bf B}=\nabla\times{\bf A}. 
\label{eq:Maxwell} 
\end{eqnarray} 
The supercurrent $\tilde{\bf j}_{\rm s}
=(\tilde{j}_{{\rm s},x},\tilde{j}_{{\rm s},y})
\propto(\partial \tilde{f}/\partial A_x, \partial \tilde{f}/\partial A_y)$ 
is given by 
$
\tilde{j}_{{\rm s},x}
={\rm Re} [ 
 \eta_+^\ast (q_x \eta_+) +  \eta_-^\ast (q_x \eta_-) 
+ C_2 \{ \eta_+^\ast (q_- \eta_-)+  \eta_-^\ast (q_+ \eta_+) \} 
+ C_3 \{ \eta_+^\ast (q_+ \eta_-)+  \eta_-^\ast (q_- \eta_+) \} 
]
$, 
$
\tilde{j}_{{\rm s},y}
={\rm Re}[ 
 \eta_+^\ast (q_y \eta_+) +  \eta_-^\ast (q_y \eta_-) 
-i C_2 \{ \eta_+^\ast (q_- \eta_-) - \eta_-^\ast (q_+ \eta_+) \} 
+i C_3 \{ \eta_+^\ast (q_+ \eta_-) - \eta_-^\ast (q_- \eta_+) \} 
]
$. 
The length, field, and time are, respectively, scaled by 
the coherence length $\xi_0$, $H_{{\rm c}2,0}=\phi_0/2\pi \xi_0^2$, and 
$t_0=4 \pi \xi_0^2 \kappa^2 \sigma / c^2$ with the 
normal state conductivity $\sigma$.\cite{KatoEnomotoC,KatoEnomoto}
However, we here scale $\eta_\pm$ by $\eta_0$ instead of 
$\eta_0(T)=\eta_0(1-T/T_{{\rm c}})^{1/2}$. 
$\eta_0$ is a uniform solution of $\eta_+$ when 
$\eta_-=0$ and $T=0$. 
The calculations are performed in a two-dimensional rectangular area 
with a size $200 \xi_0 \times 100 \xi_0$. 
Outside the open boundary, we set $\eta_+=\eta_-=0$ 
and $B({\bf r})=H$ with an applied field $H$. 
We set the GL parameter $\kappa=2.7$. 

\begin{figure*} [tbh] 
\includegraphics[width=13.0cm]{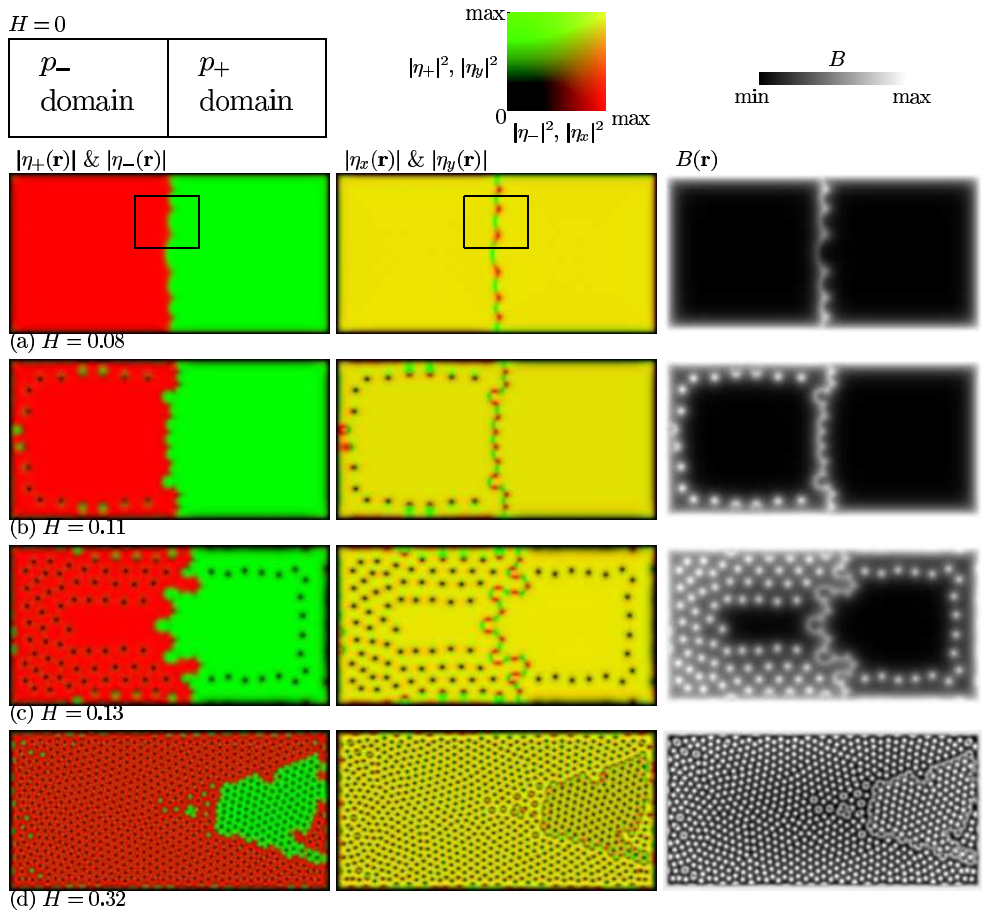} 
\caption{
(Color) 
Magnetization process for the multi-domain state 
in a $ 200 \xi_0 \times 100 \xi_0$ area with the open boundary. 
We start from the zero-field state where 
the left (right) hand side half region is a $p_+$ ($p_-$) domain. 
The applied field $H$ at the boundary is gradually increased. 
(a) $H/H_{{\rm c}2,0}=0.08$, (b) 0.11, (c) 0.13 and (d) 0.32. 
The left panels show the color density plots of 
$|\eta_+({\bf r})|$ and $|\eta_-({\bf r})|$. 
The center panels are for $|\eta_x({\bf r})|$ and $|\eta_y({\bf r})|$. 
The right panels present the internal field distribution $B({\bf r})$. 
} 
\label{fig:f1}
\end{figure*} 

Our calculation of the magnetization process for the multi-domain state 
is shown in Fig. \ref{fig:f1}. 
At a zero field, we prepare the state
where the right-hand side half region is a $p_+$ state 
($|\eta_+| \sim 1$, $\eta_- =0$),
and the left-hand side half region is a $p_-$ state 
($|\eta_-| \sim 1$, $\eta_+ =0$).
A straight domain wall appears at the center between 
the $p_+$ and $p_-$ domains. 
The relative phase of $\eta_+$ and $\eta_-$ is $\pi$,
which minimizes the free energy in our case. 
We increase $H$ gradually with a slow rate 
$\delta H/\delta t=5 \times 10^{-6}$.
The left panels in Fig. \ref{fig:f1} show the color-density plot of
$|\eta_+({\bf r})|$ and $|\eta_-({\bf r})|$.
A green (red) region indicates the $p_+$ ($p_-$) domain. 
The center panels show $|\eta_x({\bf r})|$ and $|\eta_y({\bf r})|$,
when we define the $p_x$ component $\eta_x$ and the $p_y$ component $\eta_y$ 
as $\Delta({\bf r},{\bf p})
=\eta_x({\bf r})\phi_x({\bf p})+\eta_y({\bf r})\phi_y({\bf p})$
with $\phi_x=( \phi_+ + \phi_- )/2  \sim p_x$
and  $\phi_y=( \phi_+ - \phi_- )/2i \sim p_y$. 
Yellow region indicates that  
$|\eta_x({\bf r})| \sim |\eta_y({\bf r})| \sim 1$ 
in the $p_+$ or $p_-$ domains. 
The right panels represent the internal field distribution $B({\bf r})$,
which can be observed directly.

\begin{figure} [tbh] 
\vspace{-0.5cm} 
\includegraphics[width=7.5cm]{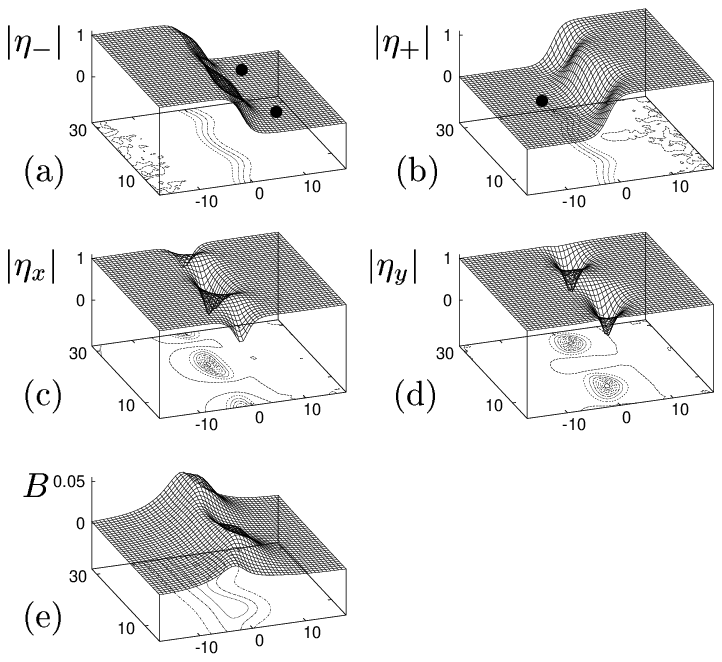} 
\vspace{-0.5cm} 
\caption{
Vortex sheet structure at $H/H_{{\rm c}2,0}=0.08$ 
in the enclosed area in Fig. 1(a). 
(a) $|\eta_-({\bf r})|$, 
(b) $|\eta_+({\bf r})|$,  
(c) $|\eta_x({\bf r})|$, 
(d) $|\eta_y({\bf r})|$, and  
(e) $B({\bf r})$ are presented. 
Solid circles in (a) and (b) indicate the position of a phase winding 
$2\pi$ for $\eta_-({\bf r})$ and $\eta_+({\bf r})$, respectively. 
} 
\label{fig:f2}
\end{figure} 

At low fields in the Meissner state [Fig. \ref{fig:f1}(a)],
magnetic fields penetrate inside through the domain wall,
forming the vortex sheet structure. 
A straight domain wall at $H=0$ begins to meander  
by the penetration of vortices at finite fields. 
To see the vortex sheet structure, we show the spatial structure 
in Fig. \ref{fig:f2}, magnifying the enclosed area in Fig. \ref{fig:f1}(a).
In the amplitude $|\eta_-({\bf r})|$ and $|\eta_+({\bf r})|$ 
in Figs. \ref{fig:f2}(a) and \ref{fig:f2}(b), we do not see the singularity 
of the vortex center. 
It is because the singularity points with the phase winding $2 \pi$ 
of vortices are at the opposite region across the domain wall, 
where the amplitude of the order parameter is well suppressed. 
That is, the winding center of $\eta_+$ ($\eta_-$) is located 
in the $p_-$ ($p_+$) domain, 
as shown by a solid circle in Fig. \ref{fig:f2}(b) (Fig. \ref{fig:f2}(a)). 
Therefore, vortices at the domain wall are core-less vortices. 

When we consider $\eta_x$ and $\eta_y$ instead of $\eta_+$ and $\eta_-$, 
as shown in the center panels in Fig. \ref{fig:f1}, 
the core-less vortices of the vortex sheet are seen 
as different structures. 
At a zero field, the domain wall is presented as a green line 
in the color density plot, 
since $\eta_y \sim 1$ and $\eta_x \sim 0$ due to the sign change 
of $\eta_x$ at the domain wall. 
When the magnetic field penetrates, 
as shown in the center panel of Fig. \ref{fig:f1}(a) or in 
Figs. \ref{fig:f2}(c) and \ref{fig:f2}(d), 
vortices of $\eta_y$ (red circle) enter along the domain wall 
from the boundary. 
These $\eta_y$ vortices are located slightly at the $p_+$ domain side 
in the domain wall region. 
The green line between the $\eta_y$ vortices changes to the vortex of 
$\eta_x$ (green circle), 
when an inter-vortices distance of the $\eta_y$ vortices becomes short 
with increasing $H$. 
That is, the order parameters $\eta_x$ and $\eta_y$ have 
vortex cores with a winding $2 \pi$ at different positions.  
This vortex sheet structure with the $\eta_x$ and $\eta_y$ vortices 
alternately aligning 
along the domain wall is the same vortex sheet structure reported 
in our previous work.\cite{Matsunaga,Matsunaga2}  
The $B({\bf r})$ distribution has a ridge along the domain wall, 
and has a peak at the $\eta_y$ vortices, 
as shown in the right panel of Fig. \ref{fig:f1}(a) or in 
Fig. \ref{fig:f2}(e). 

With increasing $H$, first, vortices penetrate inside only 
in the $p_-$ domain [Fig. \ref{fig:f1}(b)]. 
Later at higher field, vortices penetrate also in the $p_+$ domain 
[Fig. \ref{fig:f1}(c)]. 
These indicate that the lower critical field $H_{c1}$ in the $p_-$ domain 
is lower than that in the $p_+$ domain.  
This corresponds to the fact that the upper critical field 
$H_{c2}$ in the $p_-$ state is higher than that of 
the $p_+$ state.\cite{IchiokaP} 
The amplitude of the induced $p_+$ component around vortices in the 
$p_-$ state is larger than that of the induced $p_-$ component in the 
$p_+$ state.\cite{IchiokaP}  
Since $H_{c1}$ is related to the creation energy of 
the vortex core, therefore, the $p_-$ state with smaller creation energy of 
the vortex core has smaller $H_{c1}$ compared to the $p_+$ state.\cite{Heeb}  
In Figs. \ref{fig:f1}(b) and \ref{fig:f1}(c), we see vortices 
at the boundary region. 
This is because the penetrating magnetic field decreases towards 
inside of the superconductor in the length scale of the penetration depth. 
Near $H_{{\rm c}1}$ when $H$ increases, vortices first appear 
at the boundary region where locally $B({\bf r})>H_{{\rm c}1}$.

With increasing the number of vortices along the domain wall, 
the domain wall line moves meanderingly. 
With further increasing $H$, 
the domain wall moves so that the area of the $p_+$ domain shrinks 
as is seen in Fig. \ref{fig:f1}(d), 
where  the vortex sheet structure still appears along 
the meandering domain wall. 
Finally at enough high fields, the $p_+$ domain vanishes, 
and the single $p_-$ domain state is realized. 

Around the vortex cores in the $p_+$ domain and the $p_-$ domain, 
the opposite chiral component is induced as discussed 
in previous works.~\cite{IchiokaP,Heeb,Takigawa}  
We also observe some double-winding $4 \pi$ vortex near 
the boundary regions in the $p_-$ domain. 
They are seen as large green circles in the red region 
in the color density plots of $|\eta_+|$ and $|\eta_-|$ 
in Figs. \ref{fig:f1}(b)-\ref{fig:f1}(d), 
since the vortex core of the double-winding vortex in the $\eta_-$ component 
is filled by the induced $\eta_+$ component with a winding $0$. 
The population of the double-winding vortex increases 
with increasing $C_2$, which is the strength of the gradient 
coupling between the $p_+$ and $p_-$ components. 
In the $B({\bf r})$ distribution, the double-winding vortex has 
a local minimum of $B({\bf r})$ at the vortex center 
within the enhanced $B({\bf r})$ region 
around the vortex core, due to the induced $\eta_+$ component. 
Because this $B({\bf r})$ structure is not consistent to that 
observed by SANS,\cite{Kealey}  however, 
most of the vortices in ${\rm Sr_2RuO_4}$ 
are not double-winding vortices. 

\begin{figure} [tbh] 
\includegraphics[width=8.5cm]{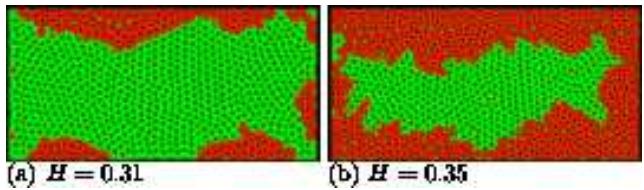} 
\caption{
(Color) 
Magnetization process in the single domain case of a $p_+$ state at $H=0$. 
Color density plots of $|\eta_+({\bf r})|$ and $|\eta_-({\bf r})|$ 
are presented at (a) $H/H_{{\rm c}2,0}=0.31$, and (b) 0.35.
} 
\label{fig:f3}
\end{figure} 

We also examine the single domain case of a $p_+$ state at $H=0$. 
Since the $p_-$ state has lower free energy in the vortex state, 
the $p_+$ domain has to change to the $p_-$ domain by applying fields.  
With increasing $H$, vortices penetrate inside in the single 
$p_+$ domain at $H \sim 0.11$. 
At a higher field $H \sim 0.25$, 
some small $p_-$ domains are created at the boundary. 
With further increasing $H$, the area of the $p_-$ domain extends 
inside, as shown in Fig. \ref{fig:f3}. 
Finally, the $p_+$ domain shrinks to vanish also in this case, 
realizing the single $p_-$ domain state.

\begin{figure} [htb] 
\includegraphics[width=5.5cm]{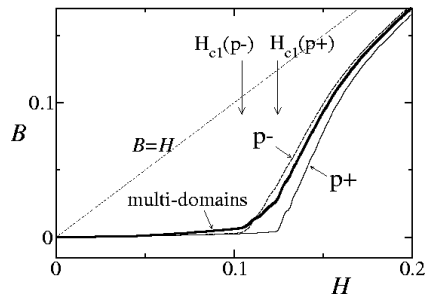} 
\caption{
Magnetization curve at low fields. 
Averaged flux density $B$ is plotted 
as a function of an external field $H$. 
We normalize $H$ and $B$ by $H_{{\rm c}2,0}$. 
When the spatial average $B$ is calculated from $B({\bf r})$, 
we exclude the boundary region of the width $10 \xi_0$ 
not to consider the surface effect at the sample boundary. 
A thin solid line is for the multi-domain case shown in Fig. 1. 
A thick solid (dashed) line is for the single domain case when 
we start from a single $p_+$ ($p_-$) domain at $H=0$. 
The dash-dotted line shows $B=H$. 
} 
\label{fig:f4}
\end{figure} 

Lastly, we compare the magnetization curves 
in Fig. \ref{fig:f4} for  three cases; 
(i) 
the multi-domain case when the $p_+$ domain and the $p_-$ domain coexist 
at low fields 
as shown in Fig. \ref{fig:f1}, 
(ii) 
the case of the single $p_+$ domain at low fields 
as shown in Fig. \ref{fig:f3}, and   
(iii) 
the case of the single $p_-$ domain.  
In the single domain case, $B=0$ in the Meissner state at $H < H_{c1}$, 
and $B$ appears in the mixed state at $H > H_{c1}$ 
due to the penetration of vortices.
We find that $H_{c1}(p_- \ {\rm state}) < H_{c1}(p_+ \ {\rm state})$ 
also in Fig. \ref{fig:f4}. 
In the multi-domain case (thin solid line in Fig. \ref{fig:f4}), 
$B$ is small but finite even in the Meissner state, 
because the magnetic field penetrates along the domain wall. 
In this case, the slope of $B$ curve changes both at 
$H_{c1}(p_+ \ {\rm state})$ and at $H_{c1}(p_- \ {\rm state})$.

In summary, 
we performed the simulation of the magnetization process 
in a chiral $p$-wave superconductor, 
using the TDGL theory with two components of the $p_+$ and $p_-$ states. 
In the multi-domain case of the $p_+$ and $p_-$ domains, 
the magnetic field penetrates as core-less vortices 
along the domain wall, forming the vortex sheet structure, 
 even in the Meissner state. 
With increasing external fields, the domain wall meanderingly moves, 
so that the area of the $p_+$ domain shrinks. 
Then, the unstable $p_+$ domain vanishes at a high field, 
and the single domain of the stable $p_-$ state is realized. 
Even in the case of the single $p_+$ domain at a zero field, 
the $p_+$ domain changes to the single $p_-$ domain 
at a high field as in a similar way, after small $p_-$ domains 
are created at the boundary. 

Recently, anomalous internal field distribution due to the 
vortex coalescence was observed.\cite{Dolocan} 
These may be related to the intrinsic character of 
the chiral $p$-wave superconductor, such as a domain structure.


\end{document}